\documentclass{elsart}
\usepackage{epsfig}
\usepackage{amssymb}

\def\be{\begin{eqnarray}}
\def\ee{\end{eqnarray}}
\def\lsim{\mathrel{\rlap{\lower3pt\hbox{\hskip1pt$\sim$}}
     \raise1pt\hbox{$<$}}} 
\def\gsim{\mathrel{\rlap{\lower3pt\hbox{\hskip1pt$\sim$}}
     \raise1pt\hbox{$>$}}} 

\def\cal{\it}

\begin{document}


\begin{frontmatter}
\title{The Problem of Mass: Mesonic Bound States Above $T_c$}

\author[pnu]{Hong-Jo Park,}
\author[pnu,apctp]{Chang-Hwan Lee,}
\author[suny]{Gerald E. Brown}

\address[pnu]{Department of Physics, Pusan National University,
              Pusan 609-735, Korea\\ 
	      (E-mail: hongjopark@pusan.ac.kr, clee@pusan.ac.kr) }
\address[apctp]{Asia Pacific Center for Theoretical Physics, POSTECH,
Pohang 790-784, Korea}
\address[suny]{Department of Physics and Astronomy,\\
               State University of New York, Stony Brook, NY 11794, USA \\
(\small E-mail: Ellen.Popenoe@sunysb.edu)}

\renewcommand{\thefootnote}{\fnsymbol{footnote}}
\setcounter{footnote}{0}

\begin{abstract}
We discuss the problem of mass, noting that meson masses decrease
with increasing scale as the dynamically generated condensate of
``soft glue" is melted (Brown/Rho scaling). We then extend the Bielefeld LGS
(Lattice Gauge Simulation)
color singlet interaction computed for heavy quarks in a model-dependent
way by including the Amp\'ere law velocity-velocity interaction.
Parameterizing the resulting interaction in terms of 
effective strength of the potential
and including screening, we find that the masses of
$\pi, \sigma, \rho$ and $A_1$ excitations, 32 degrees of freedom
in all, go to zero (in the chiral limit) as $T\rightarrow T_c$
essentially independently of the input quark (thermal) masses
in the range of $1-2$ GeV, calculated also
in Bielefeld. We discuss other LGS which show $\bar q q$ bound 
states, which we interpret as our chirally restored mesons, for 
$T > T_c$.
\end{abstract}

\end{frontmatter}

\renewcommand{\thefootnote}{\arabic{footnote}}
\setcounter{footnote}{0}
\section{Introduction\label{intro}}

The problem of mass is one of the most fundamental in physics. We know
that mesons have masses, all but the pion (in the chiral limit)
at low densities and temperatures; i.e., at low scales.
We now have experimental evidence that meson masses decrease by
$\sim 20 \%$ as the density increases to nuclear matter density. This
incipient decrease has been seen in the STAR (STAR Collaboration)
data for the $\rho$-meson,
at a low density $\sim 0.15 n_0$, where $n_0$ is nuclear
matter density \cite{Shuryak2003}.

Data on in-medium $\omega$ photoproduction in $Nb$ measured by
the CBELSA/TAPS collaboration  shows this mass to decrease roughly
consistently with about $15\%$ for nuclear matter density 
\cite{Trnka-Metag}.
Such decrease in mass predicted by Brown and Rho \cite{BR91} was
made obvious in the work referred to above as coming from tadpoles
connecting the scalar density to the relevant vector meson, as
shown in Fig.~1, basically an extension of Walecka mean field theory.

\begin{figure}
\centerline{\epsfig{file=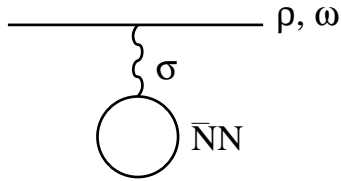,height=1.0in}}
\caption{
The decrease in vector meson mass can be simply understood as arising
from the decrease in scalar density with increasing density.
}
\label{fig1}
\end{figure}

This is well and fine for finite density, although the experiments take us up
to only $\sim n_0$. However, Harada and Yamawaki \cite{HY:PR}
carrying out a renormalization group calculation in their vector 
manifestation show that the vector mass goes to zero at a fixed
point as the temperature goes up to $T_c$ from below.

One can understand the behaviour of meson masses with temperature 
in terms of the two scales for condensed glue.

\begin{figure}
\centerline{\epsfig{file=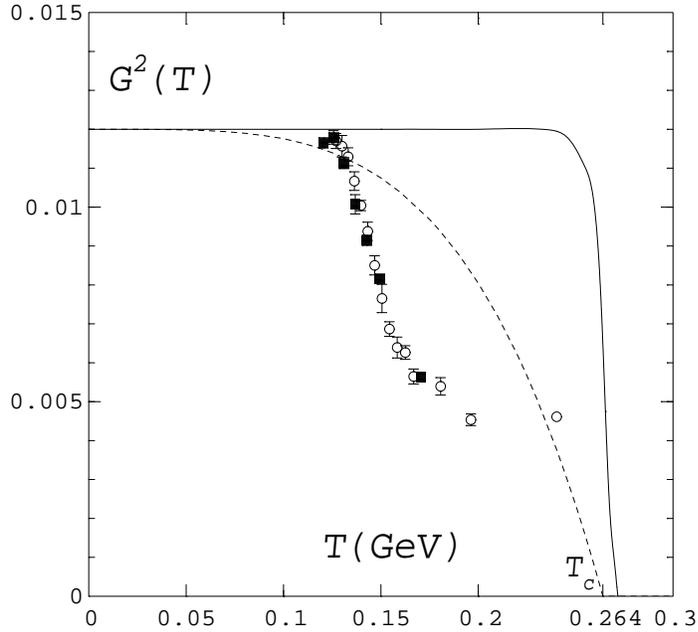,height=3.5in,
bbllx=85,bblly=240,bburx=484,bbury=611}}
\caption{Gluon condensates taken from Miller \cite{Miller}
(Fig.~2 of BGLR\protect\cite{BGLR}). The lines show the trace
anomaly for SU(3) denoted by the open circles for light quark masses
and the heavier ones by
filled circles. The $T_c$ marked in the figure is that for
quenched QCD, whereas Miller deals with unquenched QCD with
$T_c=175$ MeV.}
\label{fig2}
\end{figure}

The hard glue, or ``epoxy", is known to break scale invariance and
provide the parameter $\lambda_{QCD}$. It is an explicit
breaking by the hard glue condensate which exists up to 
$T_{\rm c, quenched} \simeq 250$ MeV (See Fig.~\ref{fig2}).

In addition to the hard glue, there is a ``soft glue", condensed below
$T_c$, which connects the quarks in mesons and provides their masses.
The soft glue provides a dynamical breaking of scale invariance, the
latter being restored at $T_c$ \cite{BR91}.

The temperature dependence of the soft glue has been calculated in
LGS by David Miller \cite{Miller} and we show it in Fig.~\ref{fig2}.
The connection of the soft glue with masses is most easily understood
in NJL (Nambu Jona-Lasinio) \cite{BrownRho2004}; it is the difference in energy obtained
by giving the quarks masses in the negative energy sea.
\be
{\rm B.E. (soft\; glue)} = 12 \int_0^\Lambda
\frac{d^3 k}{(2\pi)^3}\left[\sqrt{{\vec k}^2+{m_q^\star}^2}-|\vec k|\right]
\ee
where $m_q^\star$ is the medium dependent quark mass. As the soft glue
is ``melted" with increasing temperature, the quark mass 
$m_q^\star \rightarrow 0$ at chiral restoration,
$T_c\simeq 175 $ MeV. 
(The last point on the right in Fig.~\ref{fig2} is hard glue,
or epoxy.) The soft gluon condensate is a dynamical
breaking of scale invariance. Note that there is essentially no
change in the epoxy from the next to last point.
The soft glue holds the constituent quarks
together in mesons.

The epoxy is melted only as $T$ goes up to $T_{\rm quenched} \sim 270$ MeV.
In Shuryak's model \cite{Shuryak98}
it consists of instanton molecules which do not
break chiral symmetry.

Now the evaluation of the tadpole process of Fig.~\ref{fig1} as $T$
goes up to $T_c$ from below would involve unquenched LGS below $T_c$,
which have not yet been carried out. However, the Bielefeld group has
carried out LGS in full QCD above $T_c$. We report here on those
calculations in pure gauge, where the group has sent us
their results electronically. Their full QCD $SU(2)\times SU(2)$ results
\cite{KZ0503,Bielefeld_full2} are essentially the same as in
pure glue except for rescaling with the appropriate $T_c$'s.


\section{The $\bar q q$ Bound States at $T\gsim T_c$}
\label{sec2}

In quenched calculations Petreczky et al. \cite{Petreczky2002}
find quark and gluon (thermal) masses\footnote{These may be $\sim 30\%$
smaller in full QCD (in which they have not yet been calculated)
if rescaled with the appropriate $T_c$.
}
of $\sim 1.6$ GeV and 1.4 GeV,
respectively for $T=1.5 T_c$ where we take
$T_{\rm c, quenched} =270$ MeV. Thus, quarks and gluons are not the
thermodynamic variables of QCD in this energy range; their Boltzmann
factors would be negligible.

On the other hand, the $SU(2)\times SU(2)$ LGS transition is
second order and, as a result, the pion mass must remain zero,
in the chiral limit, as $T$ moves upwards through $T_c$.
Furthermore, the pion and sigma mesons are equivalent excitations
as chiral symmetry and helicity are restored at $T_c$,
so, in the chiral limit, these should be massless.
We have no such guiding symmetry above $T_c$ for the (equivalent)
vector and axial vector excitations, except that the vector
mass goes to zero as $T$ approaches $T_c$ from below \cite{HY:PR}.

We have shown \cite{BLR2005} that the color singlet interaction mixes
the $\pi$ and $\rho$ just above $T_c$ so strongly that one cannot
talk about linear excitations until the temperature is well
above $T_c$, the width of the excitations being as large as
the $\rho$ mass, which was taken to be $\sim 280$ MeV as arising
completely from the explicit chiral symmetry breaking.
In any case any additional mass assigned to this very nonlinear excitation
is kept small by the minimization of the Helmholtz free
energy \cite{BLR2005}. The heavy quark interaction in LGS accomplishes
only a small part of the $\pi$, $\rho$ mixing.

We have 32 degrees of freedom which all go nearly massless at $T_c$
in the chirally restored mesons $\pi, \sigma, \rho, A_1$.
This is sufficient for LGS to give the correct entropy at $T_c$.

\section{Lattice Gauge Calculations above $T_c$}
\label{sec3}

The Bielefeld group \cite{KKZP,KKPZ} 
have carried out lattice
gauge calculations to obtain the heavy quark free energy
for the region of temperatures above $T_c$.
They have given us their results for pure gauge calculations
electronically. These agree with their $SU(2)\times SU(2)$ full
QCD calculations \cite{Bielefeld_full2,Bielefeld_full} in almost all respects,
once a rescaling by the relevant temperatures is made.
We discuss in this section their results for the color singlet
(Coulomb) potentials. 

The finite temperature analog of the static potential is not known and
it is not clear whether at all it can be properly defined. Therefore
we use the internal energy as potential in the Klein-Gordon equation.
The color singlet internal energy can be derived from the free energy
obtained in lattice gauge calculations:
\be
V_{1}(r,T)=F_1 (r,T)- T \frac{\partial F_1 (r,T)}{\partial T}
\label{V1}
\ee
Above $T_{c}$, at large distance, the dominant scale of 
the effective strength of QCD potential
is the temperature, i.e. $g\approx g(T)$
for $rT \gg 1$. However, at short distances, the effective strength is 
dominated by the hard processes and 
the dominant scale is the length, i.e. $g\simeq g(r)$. 
In order to incorporate
these properties into the effective strength of the potential \cite{Felix}, 
motivated by the Debye screened perturbative results, 
we fit the color singlet free energy $F_{1}(r,T)$ as follows \cite{KKZP},
\be
F_{1}(r,T)  &=&  -\frac{g^2 (r,T)}{3\pi r} e^{-D g(r,T) T r}, \nonumber\\
g^2(r,T) &=&  \frac{A}{\log[(1/r+BT)/\Lambda]} 
\label{F1}
\ee
where A, B, and D are fitting parameters which depend on temperature. 
In principle $\Lambda$ also depends on temperature, but for 
consistency, $\Lambda=232$ MeV is used for all temperatures.\footnote{ 
Since we fitted the Bielefeld LGS data, final results are not sensitive
to the choice of different parameterizations.}
In the above expression (Eq.~(\ref{F1})) we subtracted the value of the
free energy at $r\rightarrow \infty$, namely $F_1 (r=\infty,T)$.
This is because we are only interested in the binding energy.
In Kaczmarek et al. \cite{KKZP} the authors note that ``at sufficiently
short distances the free energy agrees well with the zero temperature
heavy quark potential and thus also leads to a temperature independent
running coupling. The range of this short distance regime is temperature
dependent and reduces from $r\simeq 0.5$ fm at $T\simeq T_c$ to
$r\simeq 0.03$ fm at $T\simeq 12 T_c$."
Since the rms radius of our chirally restored mesons is $0.2$ fm at $T_c$,
it is clear that for the regime of $T$ in the neighborhood of $T_c$,
the running coupling is the same as the much studied zero temperature one.

\begin{figure}
\centerline{\epsfig{file=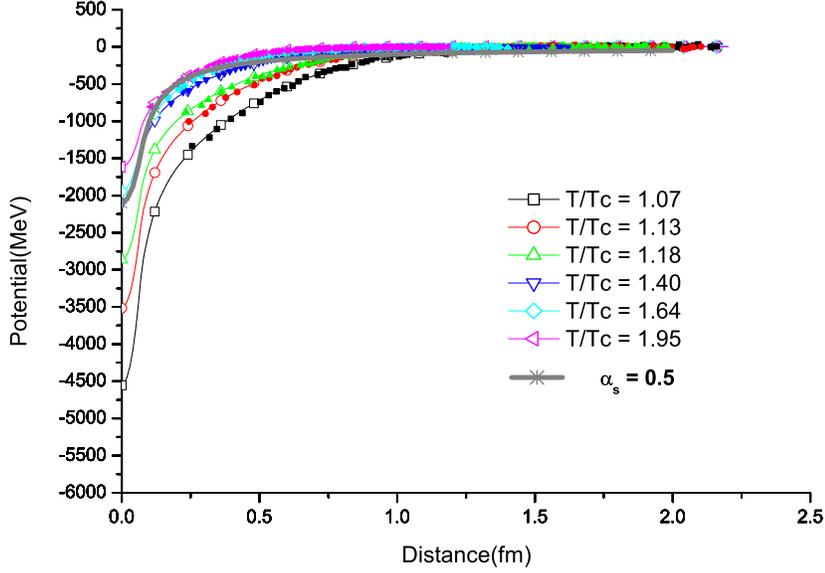,height=3.5in}}
\caption{The potential for various temperature($m_q =1.4GeV$). 
Filled symbols denote the results of lattice gauge 
simulation \cite{Kaczmarek2004b} and solid lines are 
fitted curves. 
The thick solid line is for the color-Coulomb interaction 
with constant coupling $\alpha_s=0.5$, as used
in \cite{BLRS}.
}
\label{fighj1}
\end{figure}

As in Brown et al.\cite{BLRS}, in order to enforce the
asymptotic freedom at the origin,
we used the molecule radius as
\be
R \simeq \frac{\hbar}{2m_q}.
\label{eq4}
\ee
This matches quite well the short-distance lattice regularization.
For the outside region, $r>R$, the effective potential can be determined 
using Eqs.(\ref{V1}) and (\ref{F1}) 
by fitting the results of lattice gauge simulation
summarized in Fig.~\ref{fighj1}.
In order to extend the effective potential inside the molecule radius,
the effective strength of the potential $\alpha_s^{\rm eff}$
can be defined by
\be
V(r,T) = -\frac{\alpha_s^{\rm eff}(r,T)}{r}\;\;\;\; {\rm for}\; r>R.
\label{veff}
\ee
Inside the molecular radius, following Brown et al. \cite{BLRS},
one can introduce the effective potential as
\be
V(r,T) = -\frac{\alpha_s^{\rm eff} (R,T)}{2R} 
\left(3-\frac{r^2}{R^2}\right) \;\;\;\; 
{\rm for}\;  r<R,
\ee
which is the same as the Coulomb potential for a uniform charge
distribution.

Lattice calculations from the Bielefeld group \cite{KKZP} 
are the interactions between
heavy quarks, so we need to add (model dependent) interactions
so as to use them for the light quarks making up our mesons.
Brown \cite{Brown52} showed that in a stationary state the E.M. interaction Hamiltonian
between fermions is 
\be
H_{int} = \frac{e^2}{r} ( 1- \vec{\alpha}_{1} \cdot \vec{\alpha}_{2}), 
\label{Hint}
\ee
where $\vec{\alpha}_{1,2}$ are velocity operators. 
In this paper we restrict our consideration to $T\sim T_c$. It will
turn out that the mesons are essentially massless in this region
so $\vec\alpha_1\cdot\vec\alpha_2=\pm 1$. Note that $\vec \alpha$ is essentially
the helicity.
Hence we expect for the light quark potential to be
\be
V_{\rm light}^{\rm eff} (T=T_c) &=& \left\{ \begin{array}{lll}  
2 V_{\rm heavy}^{\rm eff}
& \qquad for & \vec{\alpha}_{1} \cdot \vec{\alpha}_{2}=-1 
\\ 0 &\qquad  for& \vec{\alpha}_{1} \cdot \vec{\alpha}_{2}=+1 
\end{array} \right. 
\label{eqVeff}
\ee
since helicity is either $\pm 1$ near $T_c$.

As we shall see from the results of our calculations the factor of 2 introduced
by the magnetic interactions is precisely the factor needed to make the
(zero) masses of $\pi$ and $\sigma$ continuous across $T_c$, as dictated by
chiral symmetry (since the pion mass is protected and the $\sigma$ is
degenerate with the pion at $T_c$).

Note that the situation above $T_c$ with quarks having thermal masses is
unfamiliar to many research workers, but it is completely analogous to that
in heavy atoms where the thermal mass is simply a (repulsive) Coulomb
potential, which transforms like the 4th component of a 4-vector. Indeed,
the magnetic interaction in large $Z$ atoms has been studied by
Brenner and Brown \cite{Brenner}. Of course both Coulomb and magnetic
interactions are repulsive here. Following their Eq.~(23) it is pointed
out that the magnetic interaction is 60\% of the Coulomb for $Z=100$.
In atomic physics we do not have chiral
invariance to constrain our result, nor can we go beyond $Z\alpha=1$
with Dirac wave functions from pointlike charges, although it can go
further with our modification Eq.~(\ref{eq4}),
but the physics is the
same since we deal with Coulomb interactions and the results are in the
same ballpark. Furthermore, the Brenner and Brown results fit the measured
$K$-absorption edges of heavy atoms.
 
\begin{table}[ht]
\caption{Binding Energy due to color-Coulomb interaction $\Delta E_C$
with velocity-velocity interaction $\vec\alpha_1\cdot\vec\alpha_2=-1$, 
4-point interaction energy $\Delta E_4=G|\psi(0)|^2$, 
the radius of bound state $\sqrt{\langle r^2 \rangle}$,
reduced 4-point interaction energy $\Delta E_{\rm 4,R}$,
and the mass of bound state $M_{\rm B.S.}$. Energies and masses are in GeV,
and the radius is in fm.}
\label{tab1}
\vskip 5mm
\begin{center}
	\begin{tabular}{l|lllll}
	\hline 
$T/T_c$  & $\Delta E_C$ &  $\Delta E_4$ &$\sqrt{\langle r^2\rangle}$ 
	& $\Delta E_{\rm 4,R}$ & $M_{\rm B.S.}$ \\ 
	\hline
	&    &   $m_q =$&$1.0GeV$  & &  \\
	1.07     &  1.55  & 10.74& 0.12  & 0.039  &  0.42  \\
	1.13     &  1.33  & 5.56 & 0.16  & 0.050  &  0.62  \\
	1.18     &  1.12  & 3.13& 0.21   & 0.059  &  0.82  \\
	1.40     &  0.70  & 1.35& 0.29   & 0.070  &  1.23  \\
	1.64     &  0.56  & 1.19& 0.31   & 0.075  &  1.37  \\
	1.95     &  0.34 & 0.73&  0.37   & 0.082  &  1.57  \\
	2.61     &  0.16 & 0.53&  0.47   & 0.122  &  1.72  \\
	\hline 
	&    &   $m_q =$&$1.4GeV$  & &  \\
	1.07     & 2.03   & 21.30 &0.10   & 0.045  & 0.72  \\
	1.13     & 1.69   & 10.52 & 0.14  & 0.057  & 1.05  \\
	1.18     & 1.37   & 5.66  & 0.18   &0.065  & 1.36  \\
	1.40     & 0.85   & 2.35  & 0.24   &0.071  & 1.88  \\
	1.64     & 0.69   & 2.11  & 0.25   &0.074  & 2.04  \\
	1.95     & 0.45  & 1.26  &  0.30   &0.076  & 2.28  \\
	2.61     & 0.26  & 1.04  &  0.35   &0.095  & 2.45  \\
	\hline 
	&    &   $m_q =$&$2.0GeV$  & & \\
	1.07     & 2.68  & 44.71 & 0.08  & 0.051  & 1.27 \\
	1.13     & 2.14  & 21.03 & 0.11  & 0.064 &  1.80 \\
	1.18     & 1.68  & 10.82 & 0.15  & 0.073  & 2.25 \\
	1.40     & 1.02  & 4.24  & 0.20  & 0.075  & 2.91 \\
	1.64     & 0.85  & 3.80  & 0.21  & 0.075  & 3.08 \\
	1.95     & 0.57  & 2.22  & 0.25  & 0.073  & 3.36 \\
	2.61     & 0.38  & 1.94  & 0.27  & 0.081  & 3.54 \\
	\hline 
	\end{tabular}
\end{center}
	\end{table}

Using the effective potential enhanced by velocity-velocity interaction,
Eq.~(\ref{eqVeff}), we have estimated the color-Coulomb binding energies
and the size of bound state in Table~\ref{tab1}. 
In these estimates, we introduced the
thermal quark mass\cite{BLRS,Weldon1982} in the range of $1-2$ GeV.
This mass enters in the equation of motion as in Weldon \cite{Weldon1982}
which is different from the massless Dirac equation. The final equation
is close to the Klein-Gordon equation within 10\% error, so we solved
the Klein-Gordon equation instead of the full dispersion relation.

As noted in BLRS \cite{BLRS} we solved the Klein-Gordon equation in the
form given by Hund and Pilkuhn \cite{Pilkuhn00}
\be
\left[(\epsilon -V(r))^2 -\mu^2 - {\hat p}^2 \right] \psi(r) =0
\ee
where $\hat p$ is the momentum operator, and the reduced energy
and mass $\epsilon =(E^2-m_1^2-m_2^2)/2E$, $\mu= m_1 m_2/E$ with
$m_1=m_2=m_q$. 

In addition to the color-Coulomb interaction, there exist 4-point interactions
due to the instanton molecule interaction 
as indicated by the gluon condensate
summarized in Fig.~\ref{fig2} \cite{Miller}. In Brown et al. \cite{BGLR},
we found that the 4-point coupling constant to be
$G=3.83$ GeV$^{-2}$ above $T_c$.
In our previous approach \cite{BLRS}, we estimated that the 4-point
interaction, 
\be
\Delta E_4 = G|\psi(0)|^2,
\ee
is much stronger than the color-Coulomb interaction.
These properties are consistent with our current calculation as
summarized in Table~\ref{tab1}.
However, since the size of Coulomb bound state is now much smaller
(see Table~\ref{tab1}) than the
instanton size, the volume effect has to be properly considered. 

\begin{figure}[t]
\centerline{\epsfig{file=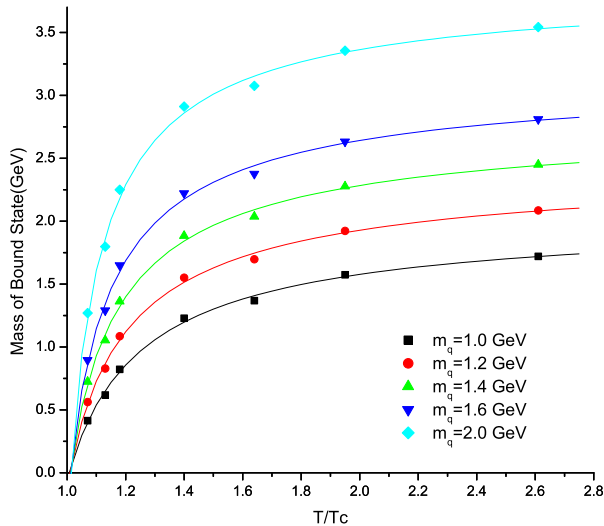,width=3.5in}}
\caption{Mass of bound states. The fitting curves show that
the mass of bound states approaches to zero as $T$ goes to $T_c$.}
\label{fig5}
\end{figure}

In this paper, we propose a new approach to the instanton molecule
interaction. The instanton molecule at $T_c$ results from the
close packing of instanton and antiinstanton in the time
direction, each of diameter 
$2/3$ fm, so the length of the instanton molecule is
\be
\frac{1}{T_c} =\frac 43 {\rm fm}.
\ee
Since the instanton radius is $1/3$ fm, the effective volume of
the instanton molecule is
\be
V_{\rm ins} = \frac 43 {\rm fm} \times \pi \left(\frac 13 {\rm fm}\right)^2
= 0.465\; {\rm fm}^3.
\ee
By considering the finite size of the Coulomb bound state,
the 4-point interaction can be reduced as
\be
\Delta E_{\rm 4,R} =\Delta E_4 
\frac{\left(\sqrt{\langle r^2\rangle}\right)^{3}}{V_{\rm ins}}.
\ee
The reduced 4-point interactions are summarized in Table~\ref{tab1}.
Note that the binding is now dominated by the color Coulomb interaction
for smaller temperatures due to the smaller size of the bound state.
The final masses of bound states are summarized in Table~\ref{tab1}
and Fig.~\ref{fig5} for different thermal quark masses.
The figure clearly shows that the mass of bound state approaches zero
as $T$ decreases to $T_c$, independently of the thermal quark masses.
The bound states in Table~\ref{tab1} include the effects of the 
velocity-velocity interaction. 
 
\begin{table}[ht]
\caption{Binding Energy due to color-Coulomb interaction $\Delta E_C$
with only heavy quark interaction 
(without velocity-velocity interaction $\vec\alpha\cdot\vec\alpha_2$).
Notations are the same as in Table~\ref{tab1}. 
}
\label{tab2}
\vskip 5mm
\begin{center}
	\begin{tabular}{l|lllll}
	\hline 
$T/T_c$  & $\Delta E_C$ &  $\Delta E_4$ &$\sqrt{\langle r^2\rangle}$ 
	& $\Delta E_{\rm 4,R}$ & $M_{\rm B.S.}$ \\ 
	\hline
	&    &   $m_q =$&$1.0GeV$  & &  \\
	1.07     &  0.87  & 1.53& 0.27  & 0.065  &  1.06  \\
	1.13     &  0.56  & 0.77& 0.34  & 0.067  &  1.38  \\
	1.18     &  0.37  & 0.43& 0.42   & 0.069  &  1.56  \\
	1.40     &  0.12  & 0.18& 0.62   & 0.090  &  1.79  \\
	1.64     &  0.04  & 0.12& 0.87   & 0.116  &  1.84  \\
	1.95     &  0.002 & 0.02&  3.27   & 0.023  &  1.98  \\
	2.61     &  $-$   & $-$ &  $-$   & $-$  &  $-$  \\
	\hline 
	&    &   $m_q =$&$1.4GeV$  & &  \\
	1.07     & 1.03   & 2.60 &0.23     & 0.071  & 1.70  \\
	1.13     & 0.66   & 1.27 & 0.29    & 0.069  & 2.07  \\
	1.18     & 0.45   & 0.69  & 0.36   & 0.067  & 2.28  \\
	1.40     & 0.18   & 0.31  & 0.49   & 0.075  & 2.55  \\
	1.64     & 0.09   & 0.25  & 0.58   & 0.102  & 2.61  \\
	1.95     & 0.03  & 0.12  &  0.90   & 0.116  & 2.66  \\
	2.61     & $-$   & $-$  &  $-$     & $-$    & $-$  \\
	\hline 
	&    &   $m_q =$&$2.0GeV$  & & \\
	1.07     & 1.22  & 4.73  & 0.20  & 0.077  & 2.70 \\
	1.13     & 0.78  & 2.20  & 0.25  & 0.072 &  3.15 \\
	1.18     & 0.54  & 1.17  & 0.30  & 0.068  & 3.39 \\
	1.40     & 0.24  & 0.54  & 0.39  & 0.067  & 3.69 \\
	1.64     & 0.15  & 0.48  & 0.42  & 0.078  & 3.77 \\
	1.95     & 0.07  & 0.28  & 0.54  & 0.094  & 3.83 \\
	2.61     & 0.01  & 0.12  & 1.17  & 0.118  & 3.87 \\
	\hline 
	\end{tabular}
\end{center}
	\end{table}



We have used Eq.~(\ref{eqVeff}) in calculating Table~\ref{tab1}, 
so our numbers are only meaningful
at and slightly above $T_c$. Higher up we should switch over to the velocity-velocity
interaction of Shuryak and Zahed \cite{SZ2004}. We know that at a higher temperature
$T = T_{\rm zero\ binding}$ the velocities will go to zero as the molecules break up.
This is the region of the ``perfect liquid"[24].
Our Table~\ref{tab1} is still useful for $T=1.4 T_c$ where LGS have found our chirally
restored mesons using the maximum entropy method \cite{Asakawa2003}.

In Table~\ref{tab2},   
we summarized the results with
only the heavy quark interactions.
One can interpolate between $\alpha_1 \cdot \alpha_2 =1$ and zero, using the
Shuryak and Zahed \cite{SZ2004} in order to determine the velocities.

\section{Lattice Evidence for the $\bar q q$ Bound States above $T_c$}
\label{sec4}

\begin{figure}
\centerline{\epsfig{file=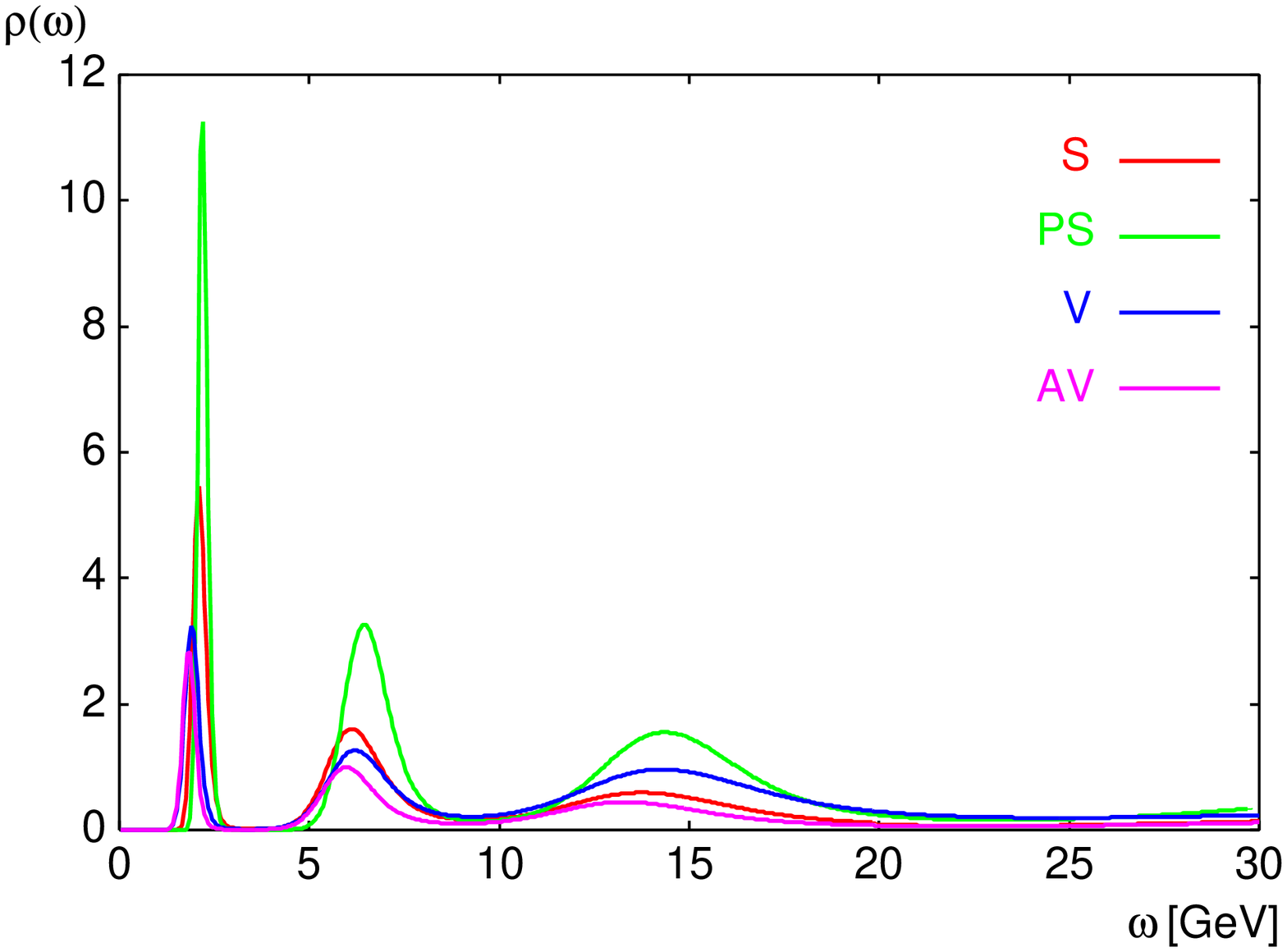,height=2in}
\epsfig{file=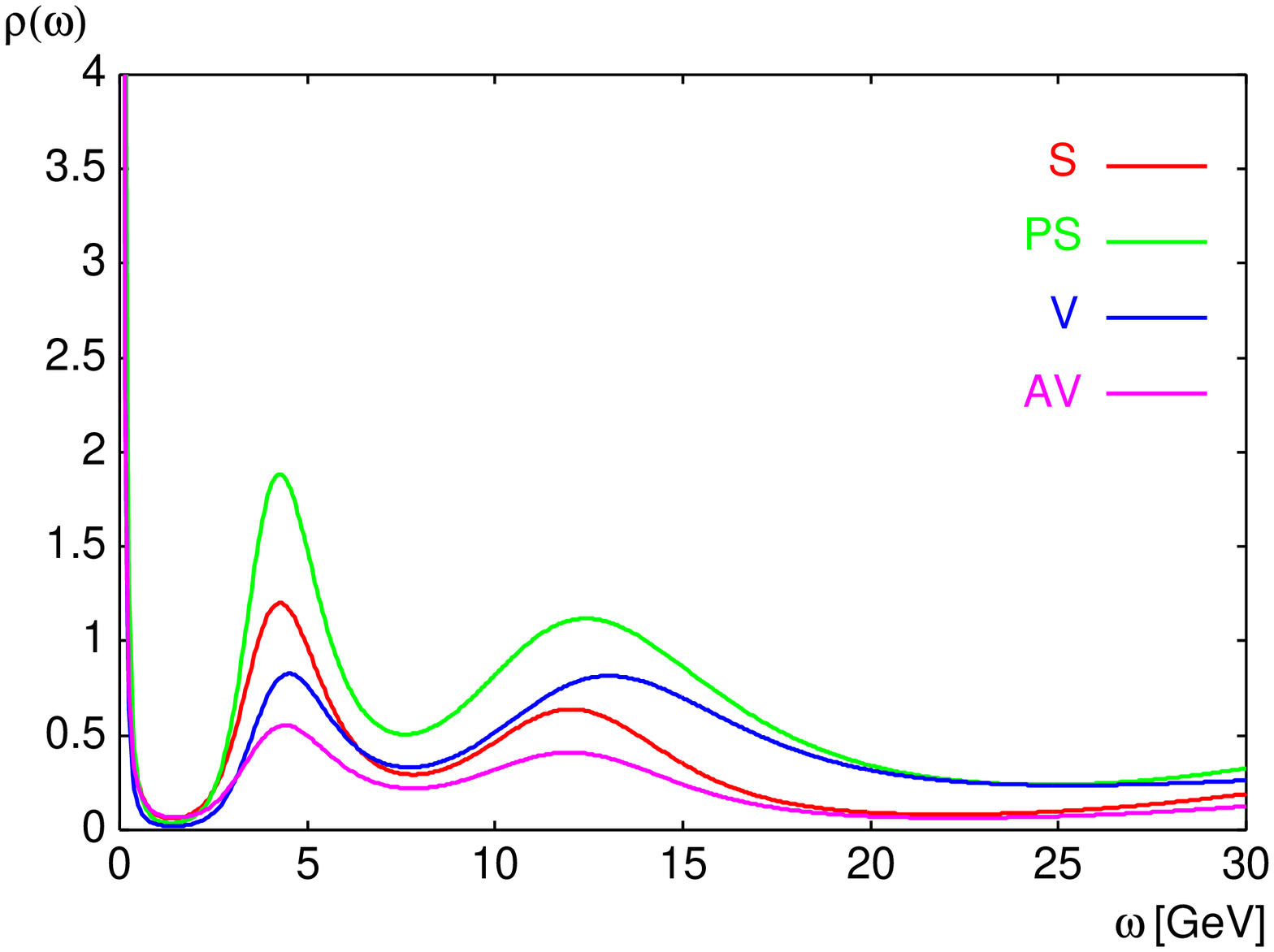,height=2in}}
\caption{Spectral functions of Asakawa et al.\cite{Asakawa2003}.
Left panel: for $N_{\tau}=54$ ($T\simeq 1.4 T_c$).
Right panel: for $N_{\tau}=40$ ($T\simeq 1.9 T_c$).}
\label{figX}
\end{figure}

In Fig.~\ref{figX} we show the spectral functions of Asakawa et al
\cite{Asakawa2003}. Similar results were obtained by
Petreczky \cite{Petreczky2004}. 
Only the peak lowest in energy at each temperature should be
considered. The higher ones are lattice artefacts.
We especially consider the 2 GeV peak at $T=1.4 T_c$. The lattice
calculations are quenched, so $T_c$ here is $\sim 270$ MeV.
Possibly all scales would come down by the factor 
($T_{\rm c, unquenched}/T_{\rm c,quenched}$) if full QCD could be
carried out.

We should remember that these vibrations were calculated in the
heavy quark approximation.

First of all, we note that there is a complete SU(4) symmetry.
In BLRS\cite{BLRS} we found no dependence of masses on isospin
and negligible dependence upon spin. Thus the degeneracy of the
S, PS, V, and AV degrees of freedom should not be surprising.

As noted in BLR \cite{BLR2005}, for temperatures not far above $T_c$,
the color singlet interaction mixes the $\rho$ and $\pi$ excitations
strongly, so the linearity that is present in the $T=1.4 T_c$ results 
would not be expected to be present at lower temperatures, even 
in the heavy quark approximation in which the effective color
singlet coupling is only half that for light quarks in our
approximation (and the square of the coupling, which comes into
the nonlinearities, only one-fourth).

In this sense, it is useful that the heavy quark interaction is
small enough so that the vibrations are roughly linear at $T= 1.4 T_c$
when calculated with it.

In any case, we see that the bound state masses increase rapidly
as $T$ moves upwards from $T_c$. However, Shuryak and Zahed \cite{SZ2004}
show that very large numbers of other excitations, mostly colored,
come in in such a way as to furnish the pressure which is lost
by the massless excitations at $T_c$ contracting out, to a large
extent, as thermodynamic variables as their mass goes up with
increase in temperature. We will not deal with the
many further excitations above $T_c$,
considered by Shuryak and Zahed \cite{SZ2004},
but simply note again that the increase
in entropy in going up to $T_c$ can be furnished by our 32 massless
mesons there.

\section{Discussion}
\label{sec5}

In BLRS\cite{BLRS} we based our construction of the $\pi$ and $\sigma$
mesons on their chiral properties; i.e., that the zero mass
(in the chiral limit) of the pion should go smoothly through $T_c$.
At $T_c$ and above the $\sigma$ is degenerate with the pion.
The symmetry in isospin followed from our Lagrangian in which
$\tau = (\vec \tau, 1)$ is a 4-vector, so that the $\rho$ and
$\omega$ were components of the same particle, also the $\sigma$
and $\pi$.
There was, however, no symmetry requirement on the mass of the
$\rho$, as $T$ went down to $T_c$ from above.
(In the vector manifestation of Harada and Yamawaki \cite{HY:PR} 
$m_\rho^\star$
must go to zero at a fixed point as $T$ goes up to $T_c$ from below.)

Spin effects are, however, small. The magnetic moment of the quark
or antiquark is \cite{BLRS}
\be
\mu_{q, \bar q} =\mp \frac{\sqrt{\alpha_s}}{p_0} 
\ee
where $p_0 \sim 2 m_{q, \bar q}$ and
$m_{q,\bar q}$ is the thermal mass. 
For our chirally restored mesons 
at $T_c$, we see 
that $\alpha_s\sim 2$,
which with Eq.~(23) of Brown et al. \cite{BLRS}, would give
$m_\rho \sim m_q/6$.

Although the confluence of curves giving the mass of the bound state
in Fig.~\ref{fig5} looks very impressive at first sight, we caution
that the data points (points from the LGS) that we have for the
lower $T/T_c$ values do not go to distances shorter than
$\sim 0.25$ fm, whereas our rms radius at $T_c$ is only
$\sim 0.5$ of this. However, Kaczmarek et al. 
(denoted as KKZP below) \cite{KKZP}
have shown that at sufficiently short distances, up to $r=0.5$ fm
at $T\simeq T_c$, the heavy quark free energy agrees well with the 
zero temperature heavy quark potential, which is essentially
Coulombic. KKZP
emphasize that in this distance range for $T_c$ (and shorter
ones for higher temperatures) processes in the QCD plasma phase
are still dominated by properties of the QCD vacuum.
The KKZP analysis suggests that ``it is more appropriate to characterize
the non-perturbative properties of the QCD plasma phase close to $T_c$
in terms of remnants of the confinement part of the QCD force
rather than a strong Coulombic force."

It's hard to separate the pure Coulombic effect from non-perturbative effects
in our approach. However, these properties are already included in our 
results because we parameterized our potential to fit the lattice results.
In our parameterization, these non-perturbative effects are 
absorbed in the effective strength of the potential, Eq.~(\ref{veff}). 
We believe that
the stronger couplings for lower temperatures towards $T_c$
indicate the stronger non-perturbative effect near $T_c$.
For small distances, $r \ll \Lambda^{-1}\sim 1$ fm,
the strength of the potential $g(r,T)$ is dominated by the radius,
close to the zero temperature limit.

We can obtain a semi quantitative picture of why the meson masses are
small at $T_c$ in the following way. KKZP show that the short
distance color averaged free energy is dominated by the singlet contribution.
This interaction is very large at large distances, as shown by the
Polyakov loops in quenched QCD (Fig.~4 of KKZP \cite{KKZP}).
Thus, overall we believe that
we can talk about Coulomb interactions (realizing that the discussion will 
not be gauge invariant) in order to provide a simple picture.

With the doubling of the effective Coulomb interaction by inclusion of the
velocity-velocity term and with the Casimir operator $4/3$, we would
obtain a (large distance)
\be
\alpha_{\rm eff} (r) = \frac{g_{\rm eff}^2}{4\pi} = \frac{16}{3}
\ee
from the Polyakov loops at $T_c$, corresponding to 
\be
g_{\rm eff}\sim 8.
\ee
Of course this implies strong coupling. 
It seems reasonable that this large
color singlet coupling is largely responsible for the large $m_q$;
indeed, rescaling the quenched 1 GeV $m_q$, which we used at $T_c$,
would give $\sim 600$ MeV for the unquenched one.

\begin{figure}[ht]
\centerline{\epsfig{file=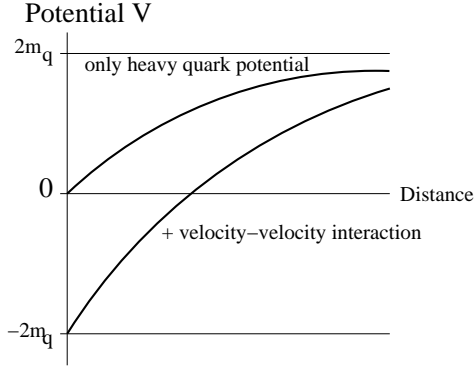,width=2.5in}}
\caption{Schematic model of the color singlet potential
and the meson binding.}
\label{fig6}
\end{figure}

In order to show why all of the curves in  Fig.~\ref{fig5} sweep
to zero, independently of $m_q$, in Fig.~\ref{fig6},
we make our picture in terms of
Coulomb energies and potentials, two repulsive Coulomb
energies $2 m_q$ being the meson energies at a large separation;
i.e., we move our zero in energy down to $-2 m_q$.

The Coulomb is attractive, but in our gauge theory, must go
to zero as the separation $r\rightarrow 0$, because of
asymptotic freedom.\footnote{As we noted earlier, 
asymptotic freedom is built into our schematic potentials \cite{BLRS}.}

Viewing the potentials as like that in the hydrogen atom, we
would expect the bound level to come down about halfway in the
Coulomb potential; i.e., the zero point energy would be about
half of the potential depth. 

With the doubling in the potential from the velocity-velocity
interaction, the molecular state comes down to $-2 m_q$.
This argument explains why the states in Fig.~\ref{fig5} all go
to zero, independently of $m_q$. Our calculations support
this picture; the potentials are approximately rescaled 
hydrogenic ones.

We see from our model how asymptotic freedom giving perturbative
interactions at short distances, can be reconciled with strong
large distance behavior. In fact, KKZP\cite{KKZP} find that
for their purpose of defining a running coupling at finite
temperature the singlet free energy is most appropriate,
as it has at short ($r\cdot {\rm max}(T,T_c) \ll 1$)
as well as large ($ r T \gg 1$, $T\gg T_c$)
distances and temperatures a simple
asymptotic behavior which is dominated by one gluon exchange; i.e.,
\be
F_1(r,T) = \left\{
\begin{array}{l}
-\frac{g^2(r)}{3\pi r},\;\;\; r\cdot {\rm max}(T, T_c) \ll 1 \\
-\frac{g^2(T)}{3\pi r} e^{-g(T) r T} 
\end{array} \right. .
\ee
In our case the meson at $T_c$ with rms radius $\sqrt{\langle r^2\rangle}
\le 0.12$ fm is well inside the $\sim 0.5$ fm Debye screening length,
so only the (upper) perturbative Coulombic expression in $F_1$ is needed.

Our essential step in going from the heavy-quark color singlet potential
to the light-quark description is to add the velocity-velocity
interaction. In BLRS\cite{BLRS} the instanton molecule interaction
played an even larger part, but with our very much smaller
Coulomb bound states here from the lattice parameters, these are
essentially negligible. The velocity-velocity potential
was derived for the interaction
of K-electrons in heavy atoms, holding for stationary states
\cite{Brown52}.
This interaction was important in reproducing the
K-absorption edges of heavy atoms \cite{Brenner},
where $v/c$ was not much less than unity.
We have shown \cite{BLR2005}
that low-mass chirally restored mesons just above $T_c$ will
explain some of the data from RHIC, which is difficult to understand
in the standard scenario, and have made predictions for other
experiments. Chiefly, however, our introduction of the
velocity-velocity dependent term allows us to separate the 64
degrees of freedom into 32 that have zero energy at $T_c$, in our model
and in the chiral limit, at $T_c$, with the other 32 lying in
mass $\sim 2 m_q$, and therefore irrelevant for the 
thermodynamics. Our approach is the only one giving exactly
the correct number (as measured in LGS) of degrees of freedom at $T_c$.

Our results using the numbers from the Bielefeld LGS essentially
confirm the scenario of Brown et al. \cite{Brown93} that chiral
restoration is reached by mesons going massless. This followed from
Brown/Rho scaling \cite{BR91}. The discussion of relevant
degrees of freedom; i.e., relevant mesons was, however, clarified
only much later in BLRS \cite{BLRS}. The 32 degrees of freedom
going massless at $T_c$ is precisely the number lattice calculations
require in order to fit the entropy increase up to $T_c$.
The pressure is given by these degrees of freedom only for a
narrow range of temperatures from $T_c$ up to $\sim 1.2 T_c$
where contributions begin from an additional large number of
binary bound states, both colorless and colored pairs
$gq, qq$ and $gg$ \cite{SZ2004}.

Although chiral restoration was discussed many years ago in
terms of chirally restored mesons, the importance of the color
Coulomb interaction was only understood recently, when lattice
calculations found charmonium to be bound up to $\sim 2 T_c$.
This statement does not take into account many earlier
published studies by Ismail Zahed, who explained to us the importance
of the color Coulomb interaction. We emphasize our indebtedness
to his work and to his tutelage.

In BLRS \cite{BLRS} we employed a schematic model for the heavy
quark color Coulomb interaction with $\alpha_s$ chosen to be
$\sim 0.5$ in order to bind charmonium up to $\sim 2 T_c$. Our
Coulomb bound states were then rather large at $T_c$,
with rms radius $\sim 1/3$ fm, roughly the size of the instanton.
Given these large bound states, the instanton molecule interaction
played an important role in bringing their energy to zero
at $T_c$.

In the neighborhood of $T_c$ the lattice color Coulomb potential we
use here is more than twice as large as the schematic one used in BLRS
\cite{BLRS}, with the results that the Coulomb bound states are so 
small that the instanton molecule interactions are negligible at $T_c$.

Aside from reproducing correctly the number of degrees of freedom
in the entropy, energy and pressure at $T_c$, our scenario should
be able to give descriptions in detail of dynamical processes,
such as the $\rho^0/\pi^-$ ratio measured by STAR Collaboration \cite{BLR2005}.
We also noted in that reference that many dileptons should be emitted
from the region just above $T_c$ where the $\rho$-mesons have low masses.
In Brown et al. \cite{BLR2005} it was incorrectly stated that the
$\rho$-mesons from the region of temperatures just above $T_c$ would
give dileptons of low invariant masses, especially where the cocktail
peaks up.
In fact, because of the $\sim 2.8 T$ thermal energy of the $\rho$
above $T_c$, the invariant dilepton masses, which contain this thermal
energy, should come in the $\sim 500$ MeV region, significantly
increasing these from the lower end of the mixed phase or from the
rhosobar, as seen in CERES.
However, since the invariant dilepton mass comes almost completely from
the thermal energy in the new RHIC component of dilepton,
the $p_\perp$ distribution should be different from those from
CERES, a much higher proportion of the new RHIC component having high 
$p_\perp$.

Our work addresses only the baryon number zero problem, since we use
LGS results and we deal only with mesons. 
We believe that it is relevant for RHIC, in which
after the baryons clear out early, the material is chiefly
composed of mesons.

\section*{Acknowledgments}
We are grateful to Felix
Zantow for the lattice gauge results from his thesis and to Olaf
Kaczmarek and Peter Petreczky. We would like to thank Mannque
Rho for many useful discussions.
G.E.B. has had many fruitful
discussions with Edward Shuryak and indebtedness to Ismail
Zahed earlier.
We also would like to thank the referees for their criticisms
which helped to improve our paper.
HJP and CHL were supported by grant No. R01-2005-000-10334-0(2005)
from the Basic Research Program of the Korea Science \& Engineering
Foundation.
GEB was supported in part by the US Department of Energy under Grant No.
DE-FG02-88ER40388. 


\appendix
\section{Appendix: Trace Anomalies, etc.}
\label{appA}

The movement towards zero mass of the 32 mesons as $T$ approaches to $T_c$
may appear somewhat miraculous, but the movement of the masses to zero
as $T$ goes up to $T_c$ from below was anticipated by 
Brown et al. \cite{Brown93}
although the strongly interacting form of matter above $T_c$ was not predicted;
i.e., it was not known that the thermal quark masses would be so large,
and that, therefore, interactions must be very strong to bring the meson
masses to zero just above $T_c$.

Adami et al. \cite{Adami} on the basis of Deng's 
lattice calculations \cite{Deng} found that about half of the glue melted
below $T_c$, and the other half above, as shown in our Fig.~\ref{fig2}.
Shuryak, in a number of papers, interpreted the glue condensed below
$T_c$ as arising chiefly from the random instantons, some of which were
melted going up to $T_c$, the others transforming into instanton molecules
which do not break chirality. 
We generally discuss the glue below $T_c$ as soft glue, that above as
epoxy, (hard glue).
It can be seen in Fig.~\ref{fig2} that the glue above $T_c$ at the last
point on the right is at the same position as the point at $T_c$; i.e.,
there has been no melting from $T_c$ until further above $T_c$.

Formally, it is more precise to talk about the dynamical breaking of
scale invariance by the soft gluon condensate below $T_c$. This is the
glue that is necessary in order to hold the quarks together to make
hadrons, and as this glue melts the hadron masses go to zero as in
Brown/Rho scaling \cite{BR91}.

The Yang-Mills equation are invariant under transformations of the full
conformal group. Here we discuss only scale transformation, the simplest
of these. A simple model in terms of scalar fields was considered by
Freund and Nambu \cite{Nambu}.
A scale transformation $\phi(x) \rightarrow \lambda \phi (\lambda x)$
on a scalar field induces the transformation on the scale invariant part
of the Lagrangian
\be
{\cal L}_{inv} (x) \rightarrow \lambda^4 {\cal L}_{inv}(\lambda x).
\ee
Now the trace anomaly in QCD can be expressed in terms of the divergence
of the dilaton current ${\cal D}_{\mu}$ whose non zero value signals
the breaking of scale invariance.
Including the effect of quark masses, the trace anomaly is given by
\be
\partial^\mu D_\mu = \Theta_\mu^\mu =\sum_q m_q \bar q q
-\frac{\beta(g)}{g} {\rm Tr} G_{\mu\nu} G^{\mu\nu}
\label{eqA1}
\ee
where $q$ and $G_{\mu\nu}$ are the usual quark and gluon fields. The second
main assumption, the one with which we are now concerned with, of Brown and
Rho \cite{BR91} is : {\it ``In order to be consistent with the scale property
of QCD, effective Lagrangian must reproduce faithfully the trace anomaly
A(1) in terms of effective fields." }

Important for us is the fact that there are two kinds of glue, soft and epoxy,
and two kinds of scale breaking, dynamical (or spontaneous) by the soft glue,
the glue that holds the quarks together to make hadrons, and the explicit
breaking, which takes place already in quenched QCD, giving the gluon
condensate the scale parameter $\lambda_{QCD}$.

Also important for Brown/Rho scaling is that the spontaneous breaking
cannot occur without the explicit breaking \cite{Nambu}.
Without the lattice, there is no preference to realize the dynamical
breaking in one region rather than another, but once $T_c$ for the
unquenched system is established, it is natural that this gives the scale
above which the soft glue has disappeared and only epoxy remains.

Consequently, there are two scales, (1) $T_c(unquenched)$: the scale
at which the soft glue has been melted; more importantly, the temperature
for chiral restoration, (2) $T_c(quenched)$: the scale at which the hard
glue has been melted.

Brown/Rho scaling is understood most simply by nuclear physicists brought
up in Walecka theory in terms of three-body interactions shown 
in Fig.~\ref{fig1}, with scalar ``tadpoles". However, these are not easily
evaluated in lattice gauge calculations, even at finite temperature,
because they would require unquenched calculations which are not yet
accessible. Furthermore, even when they are carried out, the
first ones will probably have quite large bare quark masses. On the
other hand the calculation of the gluon condensate below $T_c$, shown
in Fig.~\ref{fig2}, depends only weakly on the bare quark mass, only
through the $\sum_q m_q\bar q q$ term in the trace anomaly of 
Eq.~(\ref{eqA1}), which is small compared with the 
${\rm Tr} G_{\mu\nu} G^{\mu\nu}$ term. Therefore, it is much simpler
to study the way in which meson masses drop with temperature by lattice
calculations of the soft glue.

Of course, for the present paper concerned with $T\gsim T_c$ the soft
glue is absent and that is why the mesons we calculate have no more
scalar masses, and why their energies go to zero at $T_c$ to make
continuity with the massless $\pi$ and $\sigma$ just below $T_c$
(the $\rho$ and $A_1$ following suit because of the negligible spin-
and isospin-dependence). 

We have emphasized that our 32 chirally restored mesons give just the
number of bosonic degrees of freedom obtained from the measurement
of entropy at $T_c$ in lattice calculations. Of course, colored
excitations enter not far above $T_c$, contributing to the entropy
which is nearly constant over a substantial increase in temperature.
Whereas we refer the reader to Shuryak and Zahed \cite{SZ2004}
to understand this latter point, we wish here to emphasize the
utility of having a layer of colorless mesons just at $T_c$ and
slightly above. First of all, the interactions are very strong as
emphasized in BLR \cite{BLR2005}, and essentially inelastic so that
there is no doubt of equilibration, which is necessary in order to
define a thermodynamic variable such as temperature. Secondly, how do the
mesons know that they can get out of the fireball (be emitted) below
$T_c$, whereas they are confined above ? In our scenario they have to go
through the colorless layer of mesons at $T_c$, having deposited their
color above $T_c$ as the temperature decreases with expansion of the fireball.
In other words, our layer of colorless mesons at $T_c$ is a ``color purifier".

The way in which the mesons, massless just above $T_c$, acquire their on-shell
mass as they emerge below $T_c$ was only briefly discussed in 
BLR \cite{BLR2005}, mostly for the $\rho$ mesons.
Hadrons were not considered and the whole issue of how quantitatively
the hadrons go back on shell will require further study. Analysis of the
copious RHIC data should be useful here.


\end{document}